\documentclass[a4paper,twoside, 11pt]{article}
\usepackage{lmodern}
\usepackage{amssymb,amsmath}
\usepackage{ifxetex,ifluatex}
\usepackage{fixltx2e} 
\ifnum 0\ifxetex 1\fi\ifluatex 1\fi=0 
  \usepackage[T1]{fontenc}
  \usepackage[utf8]{inputenc}
\else 
  \usepackage{unicode-math}
  \defaultfontfeatures{Ligatures=TeX,Scale=MatchLowercase}
\fi
\IfFileExists{upquote.sty}{\usepackage{upquote}}{}
\IfFileExists{microtype.sty}{%
\usepackage[]{microtype}
\UseMicrotypeSet[protrusion]{basicmath} 
}{}
\PassOptionsToPackage{hyphens}{url} 
\usepackage[unicode=true]{hyperref}
\hypersetup{
            pdftitle={Deep Predictive Models in Interactive Music},
            pdfkeywords={Interaction; Prediction; Machine Learning; New Interfaces for Musical
Expression},
            pdfborder={0 0 0},
            breaklinks=true}
\usepackage{graphicx,grffile}
\makeatletter
\def\maxwidth{\ifdim\Gin@nat@width>\linewidth\linewidth\else\Gin@nat@width\fi}
\def\maxheight{\ifdim\Gin@nat@height>\textheight\textheight\else\Gin@nat@height\fi}
\makeatother
\setkeys{Gin}{width=\maxwidth,height=\maxheight,keepaspectratio}
\providecommand{\tightlist}{%
  \setlength{\itemsep}{0pt}\setlength{\parskip}{0pt}}
\ifx\paragraph\undefined\else
\let\oldparagraph\paragraph
\renewcommand{\paragraph}[1]{\oldparagraph{#1}\mbox{}}
\fi
\ifx\subparagraph\undefined\else
\let\oldsubparagraph\subparagraph
\renewcommand{\subparagraph}[1]{\oldsubparagraph{#1}\mbox{}}
\fi

\usepackage{fancyhdr}
\usepackage{longtable,booktabs,lscape}

\title{Deep Predictive Models in Interactive Music}

\author{Charles P. Martin\thanks{University of Oslo, Department of Informatics,
    \texttt{charlepm@ifi.uio.no},
     ORCID:
    \href{https://orcid.org/0000-0001-5683-7529}{0000-0001-5683-7529}}
   \and Kai Olav Ellefsen\thanks{University of Oslo, Department of Informatics,
    \texttt{kaiolae@ifi.uio.no},
     ORCID:
    \href{https://orcid.org/0000-0003-2466-2319}{0000-0003-2466-2319}}
   \and Jim Torresen\thanks{University of Oslo, Department of Informatics,
    \texttt{jimtoer@ifi.uio.no},
     ORCID:
    \href{https://orcid.org/0000-0003-0556-0288}{0000-0003-0556-0288}}
  } 
\providecommand{\institute}[1]{}
\institute{University of Oslo, Department of Informatics \and University of Oslo, Department of Informatics \and University of Oslo, Department of Informatics}
\date{}

\begin{document}
\maketitle
\begin{abstract}
Musical performance requires prediction to operate instruments, to
perform in groups and to improvise. In this paper, we investigate how a
number of digital musical instruments (DMIs), including two of our own,
have applied predictive machine learning models that assist users by
predicting unknown states of musical processes. We characterise these
predictions as focussed within a musical instrument, at the level of
individual performers, and between members of an ensemble. These models
can connect to existing frameworks for DMI design and have parallels in
the cognitive predictions of human musicians.

We discuss how recent advances in deep learning highlight the role of
prediction in DMIs, by allowing data-driven predictive models with a
long memory of past states. The systems we review are used to motivate
musical use-cases where prediction is a necessary component, and to
highlight a number of challenges for DMI designers seeking to apply deep
predictive models in interactive music systems of the future.
\end{abstract}

\hypertarget{sec:introduction}{%
\section{Introduction}\label{sec:introduction}}

Prediction is a well-known aspect of cognition. Humans use predictions
constantly in our everyday actions \cite{Clark:2013aa}, from the very
short-term, such as predicting how far to raise our feet to climb steps,
to complex situations such as predicting how to avoid collisions in a
busy street and, finally, to long-term planning. Prediction can be
defined as guessing unknown or future states of the world informed by
our current and past experiences. When our predictions are not accurate,
such as lifting our feet for one too many steps, the error is used as a
warning to correct our actions; in that case, the warning is the
sensation of surprise. Neuroscientists are now able to observe
prediction in action in the human brain. In particular, prediction has
been observed for visual perception \cite{Petro2016}, as well as
musical perception \cite{Ross:2016aa}. Other researchers have theorised
that prediction and expectations are key to our aesthetic appreciations
\cite{Brown2013}, and, indeed, that prediction is the fundamental basis
for intelligence \cite{Hawkins2005}.

\begin{figure}
\centering
\includegraphics{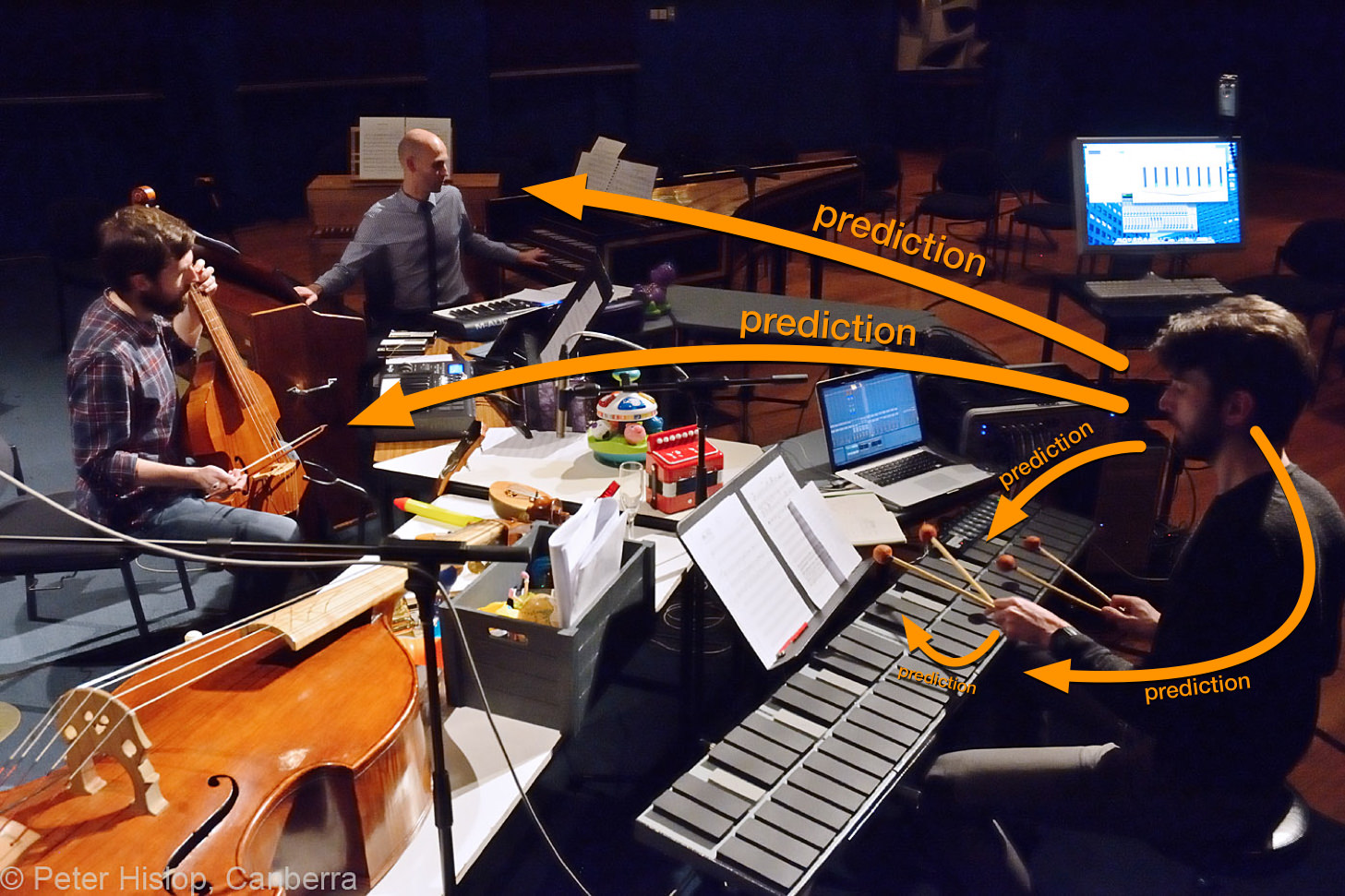}
\caption{Many types of cognitive prediction are required in musical
performance from low-level instrumental control to high-level planning
using multiple senses. Musical machine learning models can be used in
digital instruments to support performers' predictions. (Photo: © Peter
Hislop)\label{fig:prediction-performance}}
\end{figure}

Musical performance involves many layers of prediction (see Figure
\ref{fig:prediction-performance}). Skilled performers predict the sounds
produced by different instrumental gestures; they predict the musical
effect of rehearsed expressions and improvised sounds; and they predict
the musical actions of an ensemble. It may seem natural that interactive
music systems and digital musical instruments (DMIs) should incorporate
prediction to better account for the complexity of musical performance.
Brown and Gifford have noted that prediction has been only modestly
implemented in such systems \cite{Brown2013}, particularly for
incorporating proactivity into musical agents.

In contrast, we feel that many DMIs already use predictive models of
various kinds. These models are often used to generate new musical data,
manage ensemble experiences, or handle complex sensor input.
Unfortunately, the design frameworks that are often called upon to
understand these DMIs do not generally consider the role of prediction;
they tend to focus on \emph{reactive} rather than \emph{predictive}
operation.

In this paper we investigate how DMI designs using predictive models can
lead to new creative affordances for performers and DMI designers. We
draw parallels between predictive models in DMI designs and cognitive
predictions that musicians use to perform. We show how a number of
existing DMIs have applied predictive models to supplement these
cognitive predictions, extending and supporting the performer's
creativity. These systems apply various machine learning (ML) and
artificial intelligence (AI) approaches; however, we review where recent
work in deep learning has had particularly meaningful applications in
DMIs and where it could be used in future systems.

A practical contribution is that we frame two important, but usually
separate, problems in computer music---mapping and modelling---as
different sorts of predictions. Mapping refers to connecting the control
and sensing components of a musical instrument to parameters in the
sound synthesis component \cite{Hunt:2003aa}. While acoustic
instruments often have no separation between the control mechanism and
sound source (e.g., a guitar string), the separation in electronic
instruments allows the potential for many exciting and creative
mappings, but also design difficulties. Modelling refers to capturing a
representation of a musical process \cite{Dubnov:2003aa}. The model can
be used to generate further music \cite{Ames:1987aa}, or help
understand music that has been created. Both of these problems have
heuristic, as well as ML approaches. While mapping is one of the main
problems in interactive music system design, modelling is often applied
in non-real-time composition systems.

Mapping and modelling have parallels in the musical performance
predictions shown in Figure \ref{fig:prediction-performance}. Performers
learn to predict the sonic result of their control gestures; this
involves building a cognitive mapping between control and sonic output.
Performers also do higher level prediction of the notes or gestures they
play either by looking ahead in a score, or planning and selecting from
different musical possibilities in an improvisation. This clearly
involves modelling musical processes at various levels. Finally, in a
group situation, performers predict the action to sound relations and
high-level musical direction of other musicians or a conductor. This
involves both mappings and high-level models learned through experience.

By rethinking mapping and modelling as different kinds of predictions,
we can bring multiple musical applications of ML together. This exposes
some future opportunities for endowing DMIs with predictive
intelligence. It also helps to understand some of the challenges of
predictive DMIs, such as interacting in ensemble situations, and
handling temporal effects such as rhythmic, harmonic and melodic
structures.

In the next section we discuss what prediction can mean in a musical
context, review the development of musical deep learning models, and
discuss how predictive models can be incorporated into DMIs and live
musical performance environment. In Section \ref{sec:review}, we review
applications of predictive ML in two of our interactive music systems
and systems from the literature. Finally, in Section
\ref{sec:conclusions}, we examine the benefits and challenges that
predictive models can bring to DMI designers and performers.

\hypertarget{sec:predictive-models}{%
\section{Prediction and Music}\label{sec:predictive-models}}

Cognition involves many levels of prediction that we rely on for our
everyday actions \cite{Clark:2013aa}; however, it is not always clear
how prediction could be integrated into creative tools in a beneficial
way. In this section, we will discuss what prediction can mean in an
interactive system, what musical predictive models show most promise for
interactive use, and how they might fit into DMI designs and musical
performance.

\hypertarget{sec:necessary-conditions}{%
\subsection{What is a Prediction?}\label{sec:necessary-conditions}}

\begin{figure}
\centering
\includegraphics{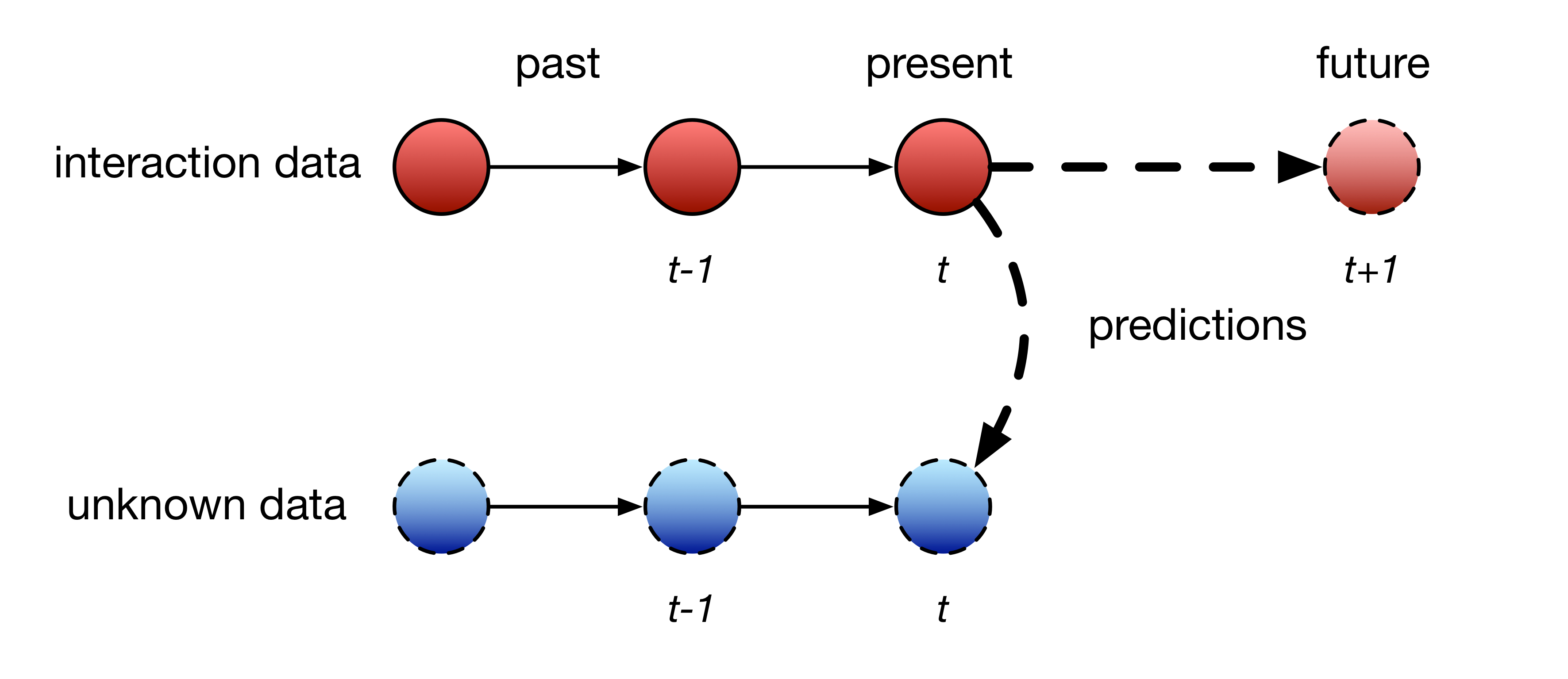}
\caption{Predictive models in an interactive context can predict future
states of a known sequence or the present state of an unknown
process.\label{fig:predictive-models}}
\end{figure}

A simple definition for prediction could be as follows: the estimation
of unknown data based on current knowledge and perceptions. This
definition encompasses the everyday understanding that prediction
relates to data in the future (e.g., weather predictions), as well as
the ML understanding of prediction as simply any unknown variable (e.g.,
image classification). In an interactive music application, perceptions
will generally consist of sensed information about the performer and
musical environment. Knowledge will consist of previous experiences
summarised in a learned model and latent variables. Unknown data can
come in two main varieties, as shown in Figure
\ref{fig:predictive-models}: future values of the sensed information, or
some different process running in parallel.

For future predictions, the sensed information may include the
performer's movements or gestures, symbolic musical data, or high-level
information about the musical context. In ML, this kind of temporal
prediction is often referred to as sequence learning \cite{Sun:2001aa}
or time series forecasting \cite{Chakraborty:1992aa}. Predictions do
not have to relate to the future. In ML, the two typical types of
prediction tasks are classification and regression, where models are
trained to predict categorical and quantitative data respectively
\cite{Hastie:2009aa}. Both of these terms are often applied when
predicting a different type of data than that given as input, without
supposing any temporal relationship. Such present predictions can have a
role in musical interaction as well; for instance, a model might predict
classifications of musical technique from gestural sensors.

\hypertarget{sec:musical-dl}{%
\subsection{Models of Musical Sequences}\label{sec:musical-dl}}

Using automatic systems to generate music is a compelling and enigmatic
idea. From the rules of counterpoint and music theory, to explorations
of indeterminacy in musical composition and performance by composers
such as John Cage or Iannis Xenakis, algorithmic composition has been
practiced for centuries. More recently, artificial neural networks
(ANNs) have been used to generate musical compositions and, now, digital
audio signals directly. Recurrent neural networks (RNNs) are often used
to generate sequences of musical notes in a one-by-one manner, where the
input is the previous note and output is the next predicted note to
occur. Mozer's CONCERT system \cite{Mozer:1994aa} is an early example
of this idea. The later introduction of gated units such as the long
short-term memory (LSTM) cell \cite{Hochreiter:1997bs} improved the
ability of such networks to learn distant dependencies. RNNs with LSTM
cells were later used by Eck and Schmidhuber to generate blues music
\cite{Eck:2002aa}. These models have a flexible ability to learn about
the temporal context in a sequence and thus mimic human cognitive
abilities for sequence learning and prediction \cite{Clegg:1998aa}.

Other popular systems for generating music use Markov models to generate
the emission probabilities of future notes based on those preceding
\cite{Ames:1989fj, Dubnov:2003aa}. The advantage of RNN models over
Markov systems is the latter requires unreasonably large transition
tables to learn distant dependencies in the data \cite{Mozer:1994aa}.
RNNs can make more ``fuzzy'' predictions, interpolating between the
training examples, rather than attempting to match them exactly
\cite{Graves:2013aa}.

The proliferation of GPU computation and large datasets has contributed
to the popularity of creative RNN models. Character-level text
generation \cite{Karpathy:2015aa}, is now well known in computational
arts. Music, too, can be represented as text and generated by an RNN
such as the ``ABC'' formatted folk songs of the FolkRNN project
\cite{Sturm:2016rz}. More complex musical forms such as polyphonic
chorales of J. S. Bach have also been modelled by RNNs; Hadjeres et al's
work on DeepBach allows such a model to be steered towards generating
voices to accompany certain melodies \cite{Hadjeres:2017aa}. RNN models
can even be combined with the rules of music theory via a reinforcement
learning tuning step described by Jaques et al.~\cite{Jaques:2017aa}.
Google's Magenta project\footnote{Magenta - Make Music and Art Using
  Machine Learning: \url{https://magenta.tensorflow.org}.} has developed
a collection of RNN models for music generation and has notably released
trained versions of several musical RNN models and used them in creative
tools and experimental interfaces.

These models learn much about the temporal structure of music, and how
melodies and harmonies can be constructed; however, there is more to
music than these aspects. Sturm et al.~\cite{Sturm:2017aa} acknowledge
as much, calling the output of FolkRNN ``transcriptions'' of (potential)
folk tunes, not tunes themselves. These transcriptions have a melody,
but musicians need to contribute their own arrangement and expression to
perform them as complete musical works.

Some recent models have begun to integrate more aspects of music into
their output, and thus produce more complete performances. Malik and
Ek's StyleNet \cite{Malik:2017aa} annotates existing musical scores
with dynamic (volume) markings. Simon and Oore's PerformanceRNN
\cite{Simon:2017aa} goes further by generating dynamics and rhythmic
expression, or rubato, simultaneously with polyphonic music. In terms of
representations of music, PerformanceRNN's output could be said to be
\emph{thicker} \cite{Davies:2005fj} than FolkRNN's thin output, because
it contains much more of the kind of information required to actually
perform a musical work.

Of course, an even thicker representation of music would be the actual
sounds of the performance. WaveNet models \cite{Oord:2016kx} can render
raw audio samples using dilated causal convolutional layers, rather than
a recurrent network, to handle temporal dependencies. These models are
capable of producing samples, the short musical sounds that can be used
in music production \cite{Engel:2017aa}, as well as translating between
different ``styles'' of music \cite{Mor:2018aa}. These models show
great promise; however, computational requirements have not been
sufficiently overcome for them to be widely explored in an interactive
context.

\hypertarget{sec:framework}{%
\subsection{Prediction in Musical Interaction}\label{sec:framework}}

\begin{figure}
\centering
\includegraphics{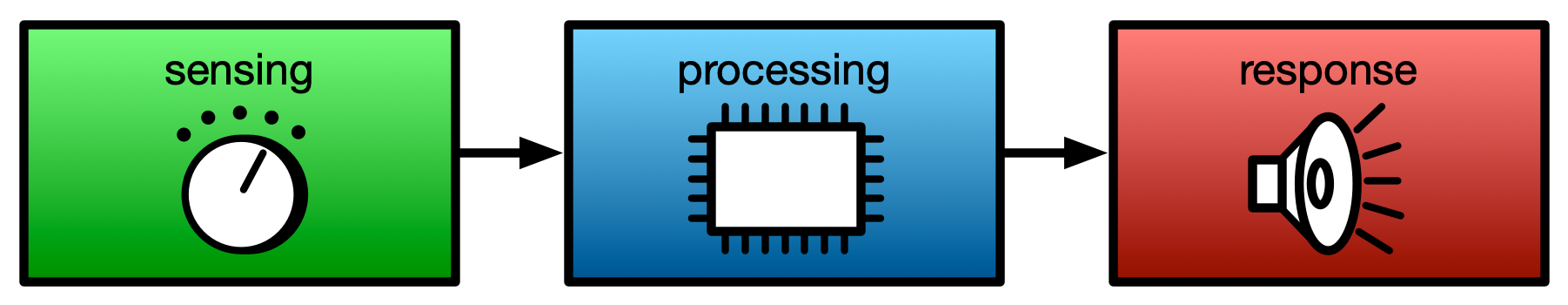}
\caption{The three-stage model for interactive music systems. A
predictive model could be considered as an extra component of this
framework.\label{fig:dmi-model}}
\end{figure}

In this section we explore where predictive models can be situated
within DMI designs and musical performance environments, and the
cognitive predictions that they could support. Interactive music systems
are often divided into three stages: sensing, processing, and response,
as shown in Figure \ref{fig:dmi-model} \cite{Rowe:1993xe}. While this
framework is simple, it provides a helpful division of concerns and has
previously been used to frame DMI designs \cite{Drummond:2009fu}
including those using ML \cite{Fiebrink:2017aa}. This framework
highlights that electronic music systems, unlike most acoustic
instruments, are modular. The sensing and response stages in particular
are often interchangeable, for instance, different interface designs
(e.g., keyboard, wind, or percussion controllers) could be used with the
same synthesiser. Complex systems may have multiple interconnected
sensors, processing stages, and responses, and might span across an
ensemble.

A predictive model can be considered as an extra component of this
framework, providing extra, or unknown, information in some part of the
DMI. Predictive models have some flexibility about the type of
information they have as input and output and can connect to this
framework at various points. Input could come from either the sensing or
response stage, and output could be directed to any stage: sensing, as
some additional generative sensed data; processing, as parameters or
adjustments to the mapping; or response, as commands for a synthesis
system.

\begin{figure}
\centering
\includegraphics{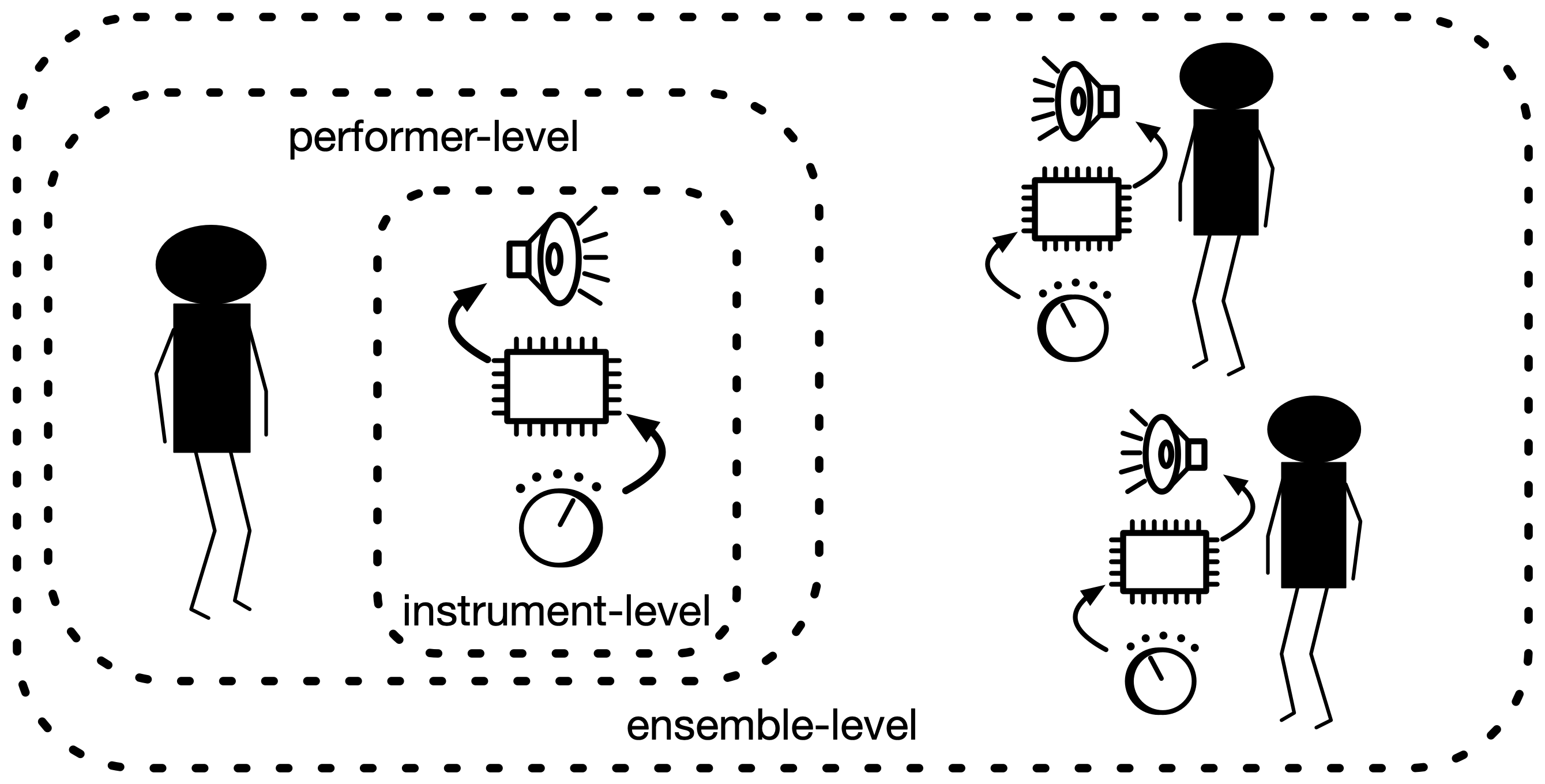}
\caption{Predictive models can be situated at various levels in musical
performance scenarios: within the instrument, at the level of the
individual performer, and between performers in a
group.\label{fig:prediction-location}}
\end{figure}

Of course, music-making involves interactions not just inside an
instrument, but between it and a performer, between the members of an
ensemble, and others such as conductors, composers and audiences. Within
this system, there are many ways to situate a predictive model that
might imitate human cognitive prediction, as discussed in Section
\ref{sec:introduction}, or make new kinds of connections. For the
purposes of our review, we will consider three levels that predictive
models could focus on within the interactive process of musical
performance. These are illustrated in Figure
\ref{fig:prediction-location} and described as follows:

\begin{enumerate}
\def\labelenumi{\arabic{enumi}.}
\tightlist
\item
  \emph{instrument-level} prediction: The model is focussed on the
  internal components of the interactive music system and generally
  predicts synthesis parameters, or aspects of the synthesis response,
  from sensor data. Thus, it replaces or supplements an instrument's
  processing stage. The model generally operates on a time-scale
  \emph{within} individual notes.
\item
  \emph{performer-level} prediction: The model predicts an individual
  performer's musical actions with their instrument over time and in
  their musical context. The actions could be represented at a low-level
  (sensor or audio data) or high-level (symbolic music or lead sheet).
  The model could predict actions that are missing or in the future, and
  is focussed on the interactions between performer, instrument, and the
  unfolding musical process. The time-scale is \emph{between} notes and
  up to large-scale musical structures.
\item
  \emph{ensemble-level} prediction: The model predicts actions of other
  members of an ensemble. This could consist of multiple performer-based
  models, or a more high-level model of interactions between ensemble
  members. The model is focussed on interactions between performers in
  the ensemble, but could operate at time-scales within or between
  notes.
\end{enumerate}

These levels for prediction correspond to typical divisions of concerns
in music performance, but could be flexible in practice given that DMIs
can be constructed with multiple connections as mentioned above. For
instance, an instrument-level model might benefit from information about
the ensemble context. Other kinds of prediction could also be considered
that include information about audiences, composers, conductors, or
other factors.

Predictive models at each level can be related to existing cognitive
predictions that a human performer uses unconsciously in order to
support or extend these functions. Instrument-level prediction relates
the movements or gestures for controlling an instrument with the pitch,
duration, volume, and timbral quality of the resulting sound, mimicking
the action-sound relationships developed when learning to play an
instrument. Performer-level prediction could allow a DMI to guess
musical actions of the performer that are missing or occur in the
future. Musicians have a similar model of musical possibilities, either
by reading ahead in a score, following a memorised piece, or improvising
new music. Many different predictions are possible at the ensemble
level; for instance, anticipating the rhythmic pulse of an ensemble,
that one musician will play a solo, or the best dynamic to enhance the
collective sound. Ensemble-level models could generate ``virtual''
ensemble members to accompany a solo performer, or predict future notes
between networked musicians to account for latency.

\hypertarget{sec:review}{%
\section{Predictive Interactive Music Designs}\label{sec:review}}

In this section, we review DMI designs that include predictive models,
including some examples of our own work (see Table
\ref{reviewed-dmis-table} for a complete listing). These examples are
divided among the three levels for prediction outlined above, and we
discuss the purpose and configuration of the predictive model in each
case. While many of these systems do not use deep learning models, they
show how predictive interaction can be incorporated into creative tools
and artistic practices.

\hypertarget{instrument-level-prediction}{%
\subsection{Instrument-level
Prediction}\label{instrument-level-prediction}}

The potential of ML models to predict the parameters of sound synthesis
systems from gestural or control input has been acknowledged since at
least the early 1990s \cite{Lee:1991aa}. One early application was Fels
and Hinton's GloveTalk II system \cite{Fels:1998fj}, where a number of
ANNs connected hand and finger sensors to a voice synthesiser. This
system was trained to produce vocal sounds from examples given by the
user. Fels and Hinton reported that GloveTalk II users needed to learn
the mappings from gesture to sound, but the ML model could also be
re-trained to better connect to gestures the user had learned so far,
thus supplementing their cognitive model for operating the system.
Predictive models at the instrument level can similarly adapt to a
user's existing model of gesture-to-sound, perhaps one learned on an
acoustic instrument, by mapping desirable sounds closer to practised
gestures. As a result, performers can learn to play new DMIs more
quickly and explore wider creative possibilities.

\hypertarget{ml-as-mapping}{%
\subsubsection{ML as mapping}\label{ml-as-mapping}}

Many artists and researchers wish to connect complex or multiple sensors
to the parameter controls of audio or computer graphics systems. As with
GloveTalk II, this can often be accomplished effectively with classical
ML models such as shallow ANNs or k-nearest neighbour classification
\cite{Altman:1992aa}. Artists have been aided in this regard by
software such as Wekinator \cite{fiebrink2009metainstrument}, that
connects such algorithms into interactive music environments, allowing
them to be trained interactively by recording examples of control data
matched to the expected output classes or parameters. In practice,
training such models on-the-fly and iteratively allows for valuable
creative exploration of their affordances and predictive power
\cite{Fiebrink:2011nx, Fiebrink:2017aa}.

Snyder's Birl \cite{jsnyder1:2014} is a series of self-contained
electronic wind instruments where continuous-valued buttons (e.g.,
capacitive sensors) are used as the control input. One iteration of the
Birl used an ANN to map between these buttons and the pitch of the
synthesised sound. This ANN was trained interactively using Wekinator,
but later implemented on a microcontroller. The advantage of the ANN
over a hand-built mapping in this case was that designed
fingering-to-note mappings could easily be learned, but the ANN also
interpolates between these fingerings (i.e., when a button is not fully
touched) and creates some, perhaps unpredictable, output for untrained
combinations. This ML approach, however, is potentially more difficult
to understand than rule-based or physical model approaches to mapping
that were also used with the Birl \cite{Snyder:2017aa}.

The use of predictive models in the processing stage is becoming more
common in interactive music designs; however, these models do not always
consider the temporal component of the data. As a result, they may not
be able to model all aspects of the musical interaction. For instance,
if a sensor can measure hand position, a non-temporal model might be
able to map the position of the hand to a response, but not the
direction of the hand's motion. Using RNNs, rather than non-recurrent
ANNs for instrument-level prediction could better account for temporal
effects in performance.

The above listed systems have all used supervised learning to generate
algorithms for instrument-level prediction, with sets of training data
provided either by a DMI designer or performer. An interesting
alternative is applied in the Self-Supervising Machine
\cite{Smith:2011aa}. In this system, real-valued input data is
segmented during performance by an adaptive resonance theory algorithm
\cite{Carpenter:1991aa}, and these examples are stored to progressively
train and re-train a shallow ANN mapping to synthesis parameters. Among
several use-cases, the model is used with input data sourced from a
touchscreen, and from the sonic features of a violin. This system allows
all learning to take place with an interactive musical performance
session; however, as the predictive model is unknown until it is
created, the performer needs to learn their own model of the DMI's
behaviour without practice, as the authors note, this ``lack of
constraints can be challenging'' \cite{Smith:2011aa}.

\hypertarget{predicting-extra-responses}{%
\subsubsection{Predicting extra
responses}\label{predicting-extra-responses}}

Many DMI designs seek to augment existing musical instruments with audio
effects, extra sounds, or visual elements. When performers literally
have their hands full, a predictive model may be able to interpret
gestural information from cameras and other sensors to control these
extra responses.

In 000000Swan's Monster \cite{Schedel:2011yg}, Wekinator's predictive
models are used to track the output from a Kinect camera and a K-Bow, a
sensor-laden bow for string instruments \cite{McMillen:2008aa}. Output
from these models are used to control triggering of audio samples,
parameters on audio effects, and computer generated visuals. The
performers provided training examples by matching demonstrations of
sensor input with desired synthesis and visual configurations in
Wekinator.

The PiaF or Piano Follower \cite{Van:2014it} is an augmented piano
system designed to track auxiliary gestures in the pianist's hands
during performances and use these to control synthesised sounds
including processing of the piano audio. The core of the system consists
of a piano keyboard connected to an audio processing system with sound
output. A Kinect depth-sensitive camera captures the position of the
performer's hands, arms, and body during the performance which is sent
to a gesture variation follower (GVF) algorithm
\cite{caramiaux2014gvf}. This temporal ML model tracks multiple
dimensions of input data to classify from a number of trained gestures.
GVF additionally provides continuous data about the speed, scale and
rotation of the gesture. This is particularly useful in a creative
interface where important expressive control, for example over timbre in
a musical instrument, could be encoded in control variations.

When operating PiaF, the performer's movements throughout a composed
performance are broken down into a sequence of gestures during a
training phase. During performances, data from the Kinect is sent to the
GVF system to determine which gesture is being performed (and thus,
which part of the performance is being played). This, and variation data
about that gesture are used to control parameters in the audio
processing part of the system. The result is a system that can enhance
the pianist's expressive options during performance.

In Monster and PiAF, the output of the predictive models were directly
tied to parameters in synthesiser and visualisation systems; however, ML
models can also be directed to more abstract, high-level,
classifications. The BRAAHMS system uses a functional near-infrared
spectroscopy (fNIRS) headset to measure the ``cognitive workload'' of a
piano performer \cite{Yuksel:2015uo} and a support vector machine (SVM)
classifier. This system adds generative harmonic lines to melodies
playing on the piano under either low- or high-workload conditions.
Ben-Asher and Leider used a naive Bayes approach on pianists' hand
movements to classify the emotional content of their performance into
six high-level categories \cite{Ben-Asher:2013aa}. These
classifications were used to drive a visualisation during performances.
Deep ANN models might be able to predict high-level information, such as
ratings of expression or rhythmic accuracy, directly from audio signals
\cite{Pati:2018ab}.

\hypertarget{performer-level-prediction}{%
\subsection{Performer-level
Prediction}\label{performer-level-prediction}}

ML models for performer-level prediction build a representation of the
notes, sounds, or actions that represent a performer's planned or
recorded interactions with an instrument. Such models are often directed
towards capturing a certain musical style \cite{Dubnov:2003aa} and
generally are configured to predict future notes on the basis of those
previously played. Performer-level models can be used in two main ways:
predicting future notes, which can be played back or compared to those
actually played, or used to analyse the music that has already been
played \cite{Conklin:2003aa}. In the following examples,
performer-level prediction is generally used to fill-in musical parts
that the performer doesn't play, or to continue when they stop.

\hypertarget{continuing-musical-interactions}{%
\subsubsection{``Continuing'' Musical
Interactions}\label{continuing-musical-interactions}}

The \emph{Continuator} is a DMI that models and imitates the style of
individual performers to ``continue'' their performances where they
leave off \cite{Pachet:2003wd}. The performer plays on a control
interface where high-level MIDI note data is sent to a synthesis module;
this MIDI data is also tracked by the Continuator. As soon as the
performer stops playing, the Continuator activates, generating new MIDI
notes in the same style as the performer's recent input and sending them
directly to the synthesiser. When the performer resumes playing, the
Continuator ceases the imitation and goes back to tracking their
performance. The temporal predictions here are generated by a variable
order Markov model that chooses from the space of various notes and
rhythms entered by the performer. This relatively simple model allows
the system to learn on-the-fly but limits the range of temporal
dependencies that can be represented.

Beatback is a model focussed on drum machine performance
\cite{Hawryshkewich:2010aa}. Similarly to the Continuator, Beatback
uses a variable order Markov model which is trained during performance
from musical material supplied by the performer. In drum machine
patterns, performers play notes on the different sounds of the acoustic
drumset: bass drum, snare, hi-hats, etc. Beatback's call-and-response
mode predicts likely continuations of the user's complete drum pattern
when they stop playing. A second ``accompaniment'' mode functions
differently, by only predicting notes for instruments that the user
leaves out. For drum machine performance, and unlike many other
instruments, predicting simultaneous musical phrases can serve as a
practical augmentation for solo performance, rather than a duet.

Deep RNN models can be applied to musical continuation in a similar way
to the Markov models. The Magenta project's AI Duet \cite{Mann:2016aa}
integrates their Melody RNN model into an interactive music system that
can run as part of a computer music environment or in a self-contained
web application. The Melody RNN attempts to predict new notes from those
in the recent past---it automatically activates during performance,
playing back its predictions in a different voice allowing the user to
engage in call-and-response style improvisation. Where the Continuator's
Markov model was trained on the performer's own playing, the Melody RNN
needs to be pre-trained on a large corpus of MIDI data. In practice, the
ability to learn from a very large corpus of data can be a significant
advantage; a novice user might provide very simple musical input and the
RNN could encourage or inspire them with more elaborate musical ideas.

The systems above operate on symbolic music as both input and output but
it is also possible to produce continuations of audio signals. OMax is a
system of agent-based predictive models designed for use in interactive
improvisation \cite{Assayag:2006aa}. This system can handle polyphonic
MIDI data as well as audio, so can be used by musicians playing acoustic
instruments. OMax allows one, or multiple, factor oracle models
\cite{Assayag:2004aa} to be trained in real-time during a musical
performance from streams of symbolic music data. The system can capture
audio signals, use pitch-tracking or some other feature analysis to
classify the signal into sequences of classes that are used to train the
predictive models, and then respond in the performer's own sound using
concatenative synthesis of the recorded signals \cite{Assayag:2006aa}.
In the future, audio feature analysis in a similar system could be
handled by a deep belief network (DBN) \cite{Hamel:2010aa} or
convolutional neural network (CNN) (e.g., \cite{Bittner:2017aa}), which
could be trained offline on larger amounts of audio data.

\hypertarget{mysong}{%
\subsubsection{MySong}\label{mysong}}

MySong is a system to automatically generate harmony accompaniments for
vocal melodies \cite{Simon:2008dp}. The predictive model takes as input
a vocal melody sung by the user and outputs a sequence of chords that
match the melody. The melody and chords can then be played back together
allowing the user to hear a piano arrangement of their performance. The
predictive model in MySong blends predictions made by a hidden Markov
model (HMM) and a simple, non-temporal model of chord probability based
on the notes that appear in each musical measure. The user is able to
tune the predictions to emphasise the HMM or melodic chord assignment,
as well as a parameter between models learned from songs divided between
major and minor modes.

The benefit of MySong's predictive model is that a user is able to hear
their vocal improvisations in the context of a full musical arrangement,
a much more complete musical work. MySong supports the user's creativity
and allows them to reflect more productively on their performances by
predicting an appropriate harmonic context. Although MySong plays back a
piano accompaniment to the melody, we categorise this model as making
performer-level predictions as the chords relate mostly to the melody
rather than the response of another performer. MySong's HMM model seems
to have been effective, but recent research suggests that bidirectional
RNNs can achieve better predictions with more diverse, and perhaps more
interesting, chord sequences \cite{Lim:2017aa}.

\hypertarget{robojam}{%
\subsubsection{RoboJam}\label{robojam}}

\begin{figure}
\centering
\includegraphics{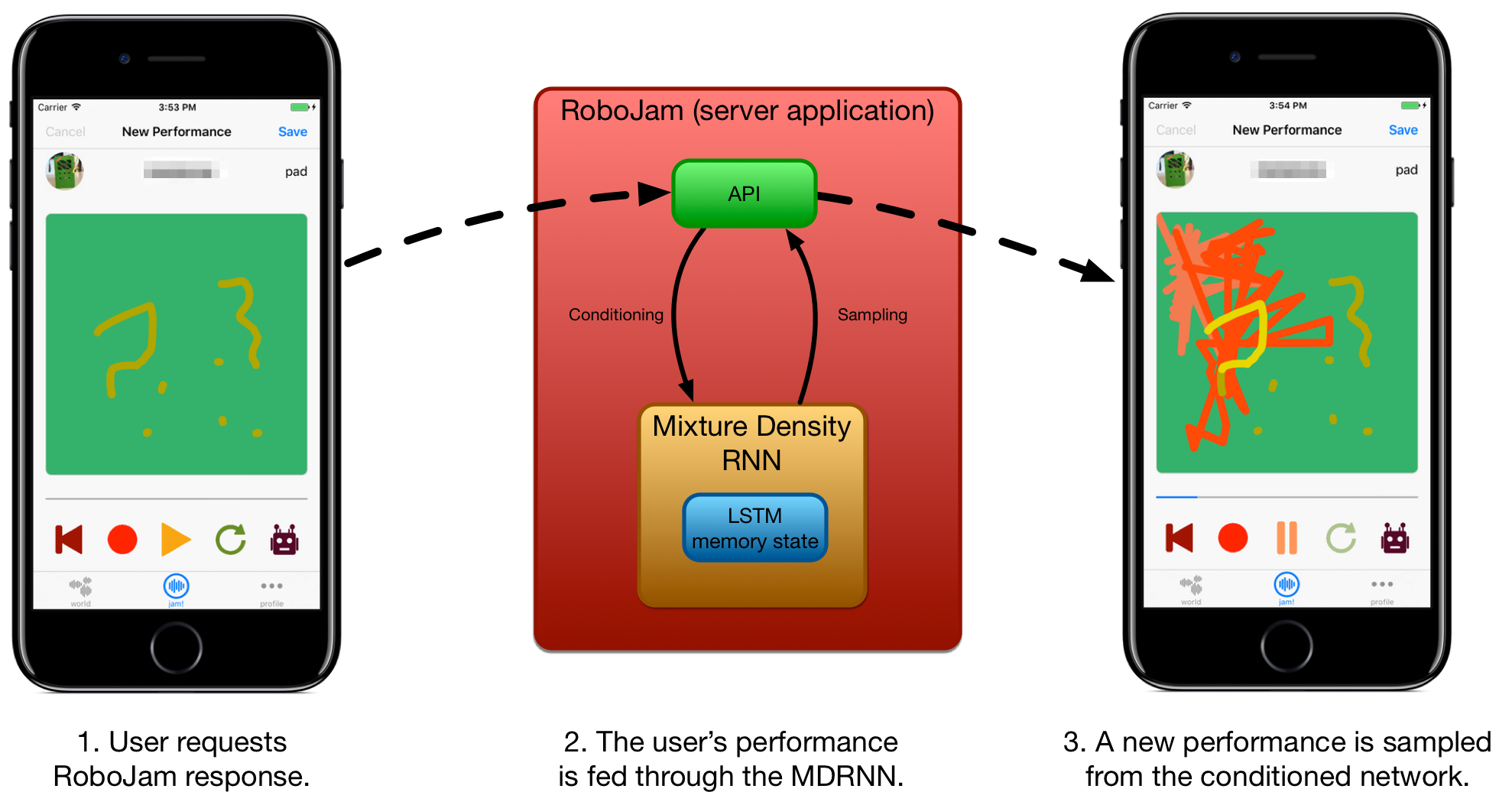}
\caption{RoboJam is a call-and-response agent for continuing touchscreen
musical performances. It uses an RNN to generate a sequence of
real-valued touch interactions after being conditioned on a user's
performance.\label{fig:robojam-interaction}}
\end{figure}

RoboJam \cite{Martin:2018ag} (shown in Figure
\ref{fig:robojam-interaction}) is a call-and-response agent developed by
the authors that uses an RNN to continue musical performances created in
a smartphone app \cite{Martin:2017ac}. RoboJam is unique in using this
RNN to model musical control data rather than musical notes. In this
way, the predictive model connects to the sensing stage of the
interactive music framework.

In this application, performers using a smartphone app collaborate
asynchronously by contributing 5-second performances to a cloud-based
music system. The short performances are created by simple mappings of
touchscreen taps and swiping to notes played by various synthesiser
instruments. RoboJam conditions an RNN on these short performances and
predicts an additional 5 seconds of control data. This predicted data is
used to play a different synthesiser and layered with the original
performance. These are musical (rather than ensemble) predictions as
they continue the performer's own control data. This system allows users
to hear more complex performances quickly, and to hear their performance
in context with different layered sounds.

RoboJam's predictive model is trained on a corpus of musical touchscreen
interactions which consist of touch locations and the time since the
previous interaction. The model uses a mixture density RNN inspired by
models of line drawings \cite{Ha:2017ab} to predict sequences of this
real-valued data. Importantly, this model is able to predict the rhythm
of interactions absolutely, rather than quantised to a set number of
steps per measure.

Since the model predicts low-level musical control data, rather than
notes, it could be said to learn how to \emph{perform} music, than how
to \emph{compose}. This arrangement means that RoboJam has access to the
whole expressive space of the touchscreen mapping and can potentially
perform very convincing responses. Since RoboJam learns to play through
the touchscreen, its performances can also be played through any of the
synthesis mappings available in the app; so, if the user performs using
a string sound, the RoboJam response might be played back with a drum
sound. While low-level learning has benefits, it comes at a cost of
difficulties in training---RoboJam's continuations are yet to be as
musically convincing as AI Duet's.

\hypertarget{ensemble-level-prediction}{%
\subsection{Ensemble-Level Prediction}\label{ensemble-level-prediction}}

In ensemble-level prediction, a model predicts the actions of other
members of a musical group. In this section we review a number of
systems in two broad categories: those that use such models to support
networked performance scenarios; and those that simulate an ensemble
from the performance of a single musician. We also discuss an example of
our own work where an RNN is used to simulate a free-improvisation
touchscreen ensemble.

\hypertarget{network-ensemble-prediction}{%
\subsubsection{Network Ensemble
Prediction}\label{network-ensemble-prediction}}

In network musical performance, groups of musicians perform together
over network connections from different physical locations
\cite{Lazzaro:2001zr}. Time delays over networks are unavoidable and
can prevent convincing performance depending on the physical distance,
system latency, and the temporal sensitivity of the music. Predictive
models between the musicians can allow information to be transmitted
ahead of actual musical events, allowing the music at each end to be
correctly synchronised.

The MalLo system accomplishes this task for percussion performances by
incorporating a predictive model into a percussion instrument
\cite{Jin:2015aa}. This model, described by Oda et
al.~\cite{Oda:2013aa}, uses computer vision techniques to track the
position of the percussionist's mallets and quadratic regression to
predict when the mallet will strike the instrument before this actually
occurs. By predicting mallet strikes, MalLo can preemptively send note
data to remote participants which is scheduled to occur in time with the
local sound. Similar systems have been implemented to predict Indian
percussion patterns \cite{Sarkar:2007aa}, and to support massed
ensemble performances using a common metronome \cite{Caceres:2008ek}.

In a related form of ensemble-level prediction, the collective behaviour
of a group is collected over a network and analysed to identify
important events in a performance. In Metatone Classifier
\cite{Martin:2015jk}, control data from a touchscreen ensemble is sent
to a central server that, first, uses a Random Forest classifier to
identify high-level gestures and, secondly, generates a Markov model to
predict whether the ensemble has collectively changed its style of
improvisation. This information is sent back to the touchscreen DMIs to
trigger changes in the individual interfaces.

\hypertarget{simulated-ensemble-prediction}{%
\subsubsection{Simulated Ensemble
Prediction}\label{simulated-ensemble-prediction}}

Individual musicians often engage in simulated ensemble experiences of
different kinds, from practice and performance with a fixed backing
track to the popular use of looping effects. With predictive models,
these experiences can be made reactive and flexible to the changing
behaviour of the performer. These applications usually include some kind
of performer-level prediction, to model the behaviour of other
individuals in an ensemble and to understand the performance of a live
soloist to provide appropriate accompaniment.

A relatively well-explored form of simulated ensemble prediction is
score following, where an algorithm tracks a performer's progression
through a known musical score to provide a synthesised accompaniment
synchronised to the soloist \cite{Dannenberg:2006aa}. The task of
tracking the performer's location is often accomplished with a hidden
Markov model where the performer's notes are the observed states and
score locations are the hidden states \cite{NIME-ScoreFollowing}. The
Orchestra in a Box system uses an HMM in this way and provides
accompaniment by playing back a time-stretched backing track
\cite{Raphael2003}.

For styles of music such as jazz, rock, and pop, a ``thick'' musical
score is usually not available, and so more advanced predictive models
are needed to create the accompaniment. In many cases, these can be
constructed using a combination of rule-based and ML systems. Biles'
\emph{GenJam} system \cite{Biles:2007aa}, for instance, uses genetic
algorithms to generate appropriate jazz-style accompaniments with
fitness determined from the rules of music theory. This system is also
able to engage in interactive improvisation with the human performer by
mutating their improvisations to create responses.

In some cases, accompaniments can be generated from a musician's own
musical material. The Reflexive Looper \cite{Pachet:2013kq} records,
manipulates and plays back audio from the musician to create an
accompaniment. Unlike a simple live looping effect that allows a
musician to record a loop and subsequent layers (typically using a pedal
interface), the Reflexive Looper uses predictive models to automatically
choose audio material to play and manipulates it to a known harmonic
progression. A support vector machine is used to classify the
performer's recent notes as either melodic, chordal or bass playing, a
generative music system then chooses appropriate backing recordings from
the two classes that are not being played. While the sound material was
generated and manipulated in real-time, the structure of the performance
in the Reflexive Looper was limited to pre-determined chord progressions
and song structures \cite{Marchini:2017aa}.

The above systems generally represent a ``virtual'' ensemble only
through sound or simple visualisation, although these musicians can also
be embodied as robots playing physical acoustic and electronic
instruments \cite{Bretan:2016aa}. For example, the marimba-playing
robot Shimon has been used in various interactive music scenarios
\cite{Bretan:2012aa}, and employs predictive models for tracking human
musicians, prediction of musical notes to play, and communication
through physical gestures. Robotic music systems require other types of
prediction to control physical movements, a focus of the SHEILA system
for imitating drum patterns \cite{Tidemann:2009aa}, but these are
beyond the scope of this paper.

\hypertarget{the-neural-touchscreen-ensemble}{%
\subsubsection{The Neural Touchscreen
Ensemble}\label{the-neural-touchscreen-ensemble}}

The Neural Touchscreen Ensemble \cite{Martin:2017ae}, a system
developed by the authors, is an RNN-driven simulation of a touchscreen
ensemble experience. A human performer plays freely improvised music on
a touchscreen and an ensemble performance is continually played back on
three RNN-controlled touchscreen devices in response. Both the human and
computer-controlled devices use a simple app that allows struck or
sustained sounds from a limited selection of notes. The performer's
touch control data is sent to the server which, using a Random Forest
classifier, predicts a high-level gestural class for the latest data
once every second. The classes come from a simple vocabulary of 9
touchscreen gestures described in previous research on iPad ensemble
performance \cite{Martin:2015jk}. The RNN uses three layers of LSTM
units and is configured to predict gestural classes in three parallel
sequences, that of the three ensemble performers. Four gestures are
taken as input---the human performer's present gesture, and the
ensemble's gesture at the last time step---and the RNN outputs the three
gestures for the ensemble at the present step. Control signals matching
the gestures can then be played back by the ensemble devices from a
corpus of performance recordings.

\begin{figure}
\centering
\includegraphics{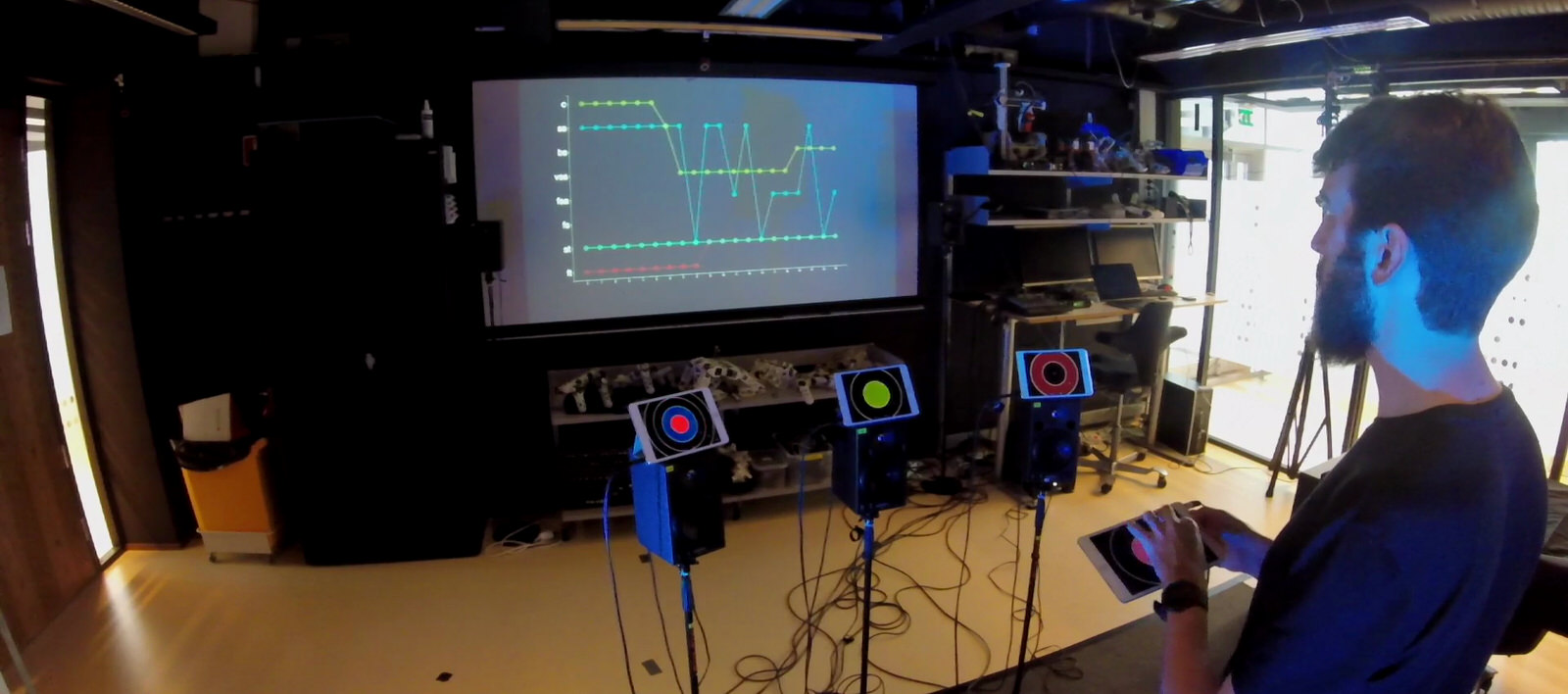}
\caption{The Neural Touchscreen Ensemble uses an RNN to predict ensemble
responses to a human performer's gestures. The system supports quartet
performances with three RNN-controlled iPads responding to one human
performer.\label{fig:neural-touchscreen-ensemble}}
\end{figure}

This system uses both an instrument-level prediction model (touchscreen
control data to gestural class) and ensemble-level prediction to predict
potential responses. The RNN ensemble model is temporal as it uses
previous experience stored in the LSTM units' state to make predictions.
Although the whole system takes touchscreen control data from the
sensing stage as both input and output, the RNN model predicts only
high-level gestures. Control data is generated from these gestures by a
touch synthesis module. In a more advanced system, touchscreen
interactions could be directly predicted as in RoboJam. The Neural
Touchscreen Ensemble's musical content---freely improvised touch
interaction---would not be easily described by music theory used in
GenJam or the Reflexive Looper above. A data-driven approach to
modelling this kind of interaction was required.

\begin{figure}
\centering
\includegraphics{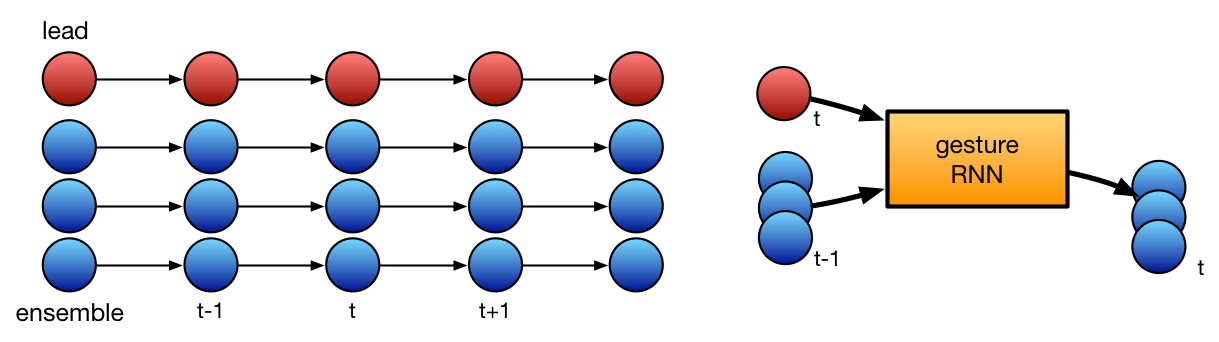}
\caption{Gesture-RNN predicts appropriate gestural motions for three
ensemble members based on present information about the human performer,
and past information about the ensemble.\label{fig:gesture-rnn}}
\end{figure}

\hypertarget{sec:conclusions}{%
\section{Conclusion}\label{sec:conclusions}}

In this paper, we have drawn parallels between predictive models in
interactive music systems with cognitive predictions involved in
performing music. We have reviewed how existing systems, including two
from our own group (RoboJam and the Neural Touchscreen Ensemble), have
implemented predictive models at the instrument-, performer-, and
ensemble-level. A variety of ML techniques have been employed, including
models that forecast future values of a known time-series, or that
predict the present value of an unknown quantity. In each case,
predicting this unknown data has allowed the systems to do more than we
would normally expect of a musical instrument. They are able to act
preemptively, to make more expressive use of the user's musical control
data, and to predict ensemble responses from artificial agents or remote
participants.

Our review demonstrates that deep learning models, in particular, have
much to offer predictive musical interaction. RNN models can learn from
large corpora of training data allowing wide musical experience to be
included in a DMI. This contrasts with Markov-based predictive models
that tend to learn only from the performer's contribution. Deep models
are flexible and can be designed to predict multiple dimensions of
related data simultaneously with the same temporal model. We took
advantage of this ability in both of our systems. In RoboJam, we were
able to predict touchscreen interactions in both 2D space and absolute
time, a novel improvement on typical step-based musical models. The
Neural Touchscreen Ensemble uses a typical RNN design, but the input and
output one-hot vectors actually encode multiple performer gestures.
Despite the interest in deep ANNs for generating symbolic music, few
interactive music systems apply these as predictive models. We suggest
that other musical deep models could be incorporated into interactive
music designs to take advantage of their flexible capacity for
data-driven prediction, and potential to generate low-level output such
as control or audio data.

Although we have discussed many interactive music systems that use ML
models, these are not often characterised in relation to cognitive
prediction. We think that this undersells the importance of prediction
in these systems and in musical performance in general. Embedding
predictive intelligence into DMIs appears to be a crucial step towards
creating interfaces that allow more expression, follow performers more
naturally, and engage more closely with ensembles. Further exploration
of their relationship with cognitive prediction could help expose ways
to use these models in music performance. In the final part of this
paper we will discuss what we see as the benefits that predictive models
can offer to DMI designers and performers, as well as some challenges
that they may face.

\hypertarget{benefits-of-prediction}{%
\subsection{Benefits of Prediction}\label{benefits-of-prediction}}

Music and sound are \emph{temporal} phenomena and yet, the widespread
framework for interactive music systems shown in Figure
\ref{fig:dmi-model} does not necessarily consider the axis of time.
Indeed, the fundamental archetype for DMIs is reactive; response
necessarily follows gesture. We think that predictive models are vitally
important to embedding a temporal axis into interactive music design. In
reality, predictive models \emph{are} used in DMI design. Considering
prediction as an essential part of interactive music design frameworks
allows these temporal models to be properly understood, examined, and
developed. This issue has gained increasing relevance in recent years
due to deep learning models enabling new insights into the difficult
problem of predicting long-term structure in music
\cite{roberts2018hierarchical}.

While traditional acoustic musical instruments are (necessarily)
reactive, their players are not. Musicians are constantly
\emph{proactive} whether anticipating a conductor's beat or introducing
a musical idea in a free jam. By embedding predictive models into DMIs,
instruments can be proactive as well, to the benefit of their users and
listeners. Indeed, in situations where reactive design is insufficient
for successful performance, such as networked ensemble performance,
predictive systems such as MalLo have been successful. We envisage that
proactive elements could be deployed much more widely in DMIs;
interfaces could change to afford upcoming musical needs as well as
respond to the users' commands.

Typical interactive music designs often include many configuration
parameters in the processing stage of their architecture. Predictive
models can be used to \emph{adapt} these parameters to meet musical
requirements of the performer, audience, or composer. In the PiaF
system, we have observed that the GVF model adapts audio processing
parameters according to the speed and size of predicted gestures.
Indeed, predictive adaptations in an interactive music system could go
much further than processing parameters. Virtual reality, touchscreen,
or haptic interfaces could be designed to adapt their complexity or
functionality according to a predicted requirement.

One of the clearest use-cases for predictive models in interactive music
design is to \emph{generate} musical data that reflects the recent style
of the user. Automatic music generation, however, can sometimes seem
like a solution in search of a problem (Who wants to listen to AI
generated music when you can play it yourself?). Both our RoboJam and
Neural Touchscreen Ensemble systems use predictive generation to enhance
solo performances. In RoboJam, response performances are generated so
that the user can hear their own work in context, while in the Neural
Touchscreen Ensemble, the actions of three RNN-controlled musicians are
generated and synthesised in real-time during the performance.

A strong motivation to continue the introduction of deep generative
models into DMIs is that the musical data of new interfaces is often
unknown and not well-modelled by music theory. Predictive RNN models,
such as that used in RoboJam, could be able to learn a wide variety of
low-level control data. Future DMIs could even use deep models with
digital audio data as input or output. These could replace multiple
parts of the three-stage DMI framework and provide multiple types of
prediction simultaneously.

\hypertarget{limitations-and-challenges}{%
\subsection{Limitations and
Challenges}\label{limitations-and-challenges}}

Adding predictive models to DMIs can present many challenges to
designers and performers. From a design perspective, it can be
challenging to develop and train ML models that are \emph{artistically
stimulating}. Environments that allow classical ML models to be trained
in near real-time (e.g.,
\cite{fiebrink2009metainstrument, amomenib:2015, Gillian:2011fu, Levy:2012aa})
assist DMI creators to experiment and evaluate the creative potential of
these models \cite{Fiebrink:2011nx}. Similarly responsive environments
for deep models are yet to appear, although the Magenta project has made
moves in this direction. As a result, the integration of RNNs and other
deep models into DMIs has been limited.

Where deep models are applied, they can present further difficulties.
Models that represent lower level data, such as the control signals in
RoboJam, tend to be more \emph{difficult to train} than symbolic music
predictors. This is partly due to larger amounts of training time and
data required in comparison to Markov systems or shallow ANNs, and is an
ongoing challenge in our research. While larger datasets can ameliorate
this issue, this may not allow a short, interactive and iterative
training process in the style of Wekinator. One workaround could be to
apply transfer learning \cite{Ng:2015aa}, where a small part of a large
pre-trained ANN is tuned using a small number of examples.

In all predictive models, the \emph{predictions are limited to the
knowledge available in the training data}. The neural touchscreen band's
RNN is trained on performance segments and not whole performances. As a
result, it can be difficult to get the simulated ensemble to start
playing, and to stop at the end of performance. This shortcoming
suggests that a few rule-based elements could be helpful, even in
data-driven models.

\emph{Understanding predictions} in a DMI can be challenging for
performers. For instrument-level prediction, this is sometimes overcome
by including the performer in the training process
\cite{Schedel:2011yg, Fels:1998fj}. For models that are trained during
performance, the performer needs to continually update their own
understanding of the model in parallel, which can become overwhelming.
DMIs that change their mappings under a musician's fingers run the risk
of frustrating rather than engaging the performer. Predictive models
that simply continue when the performer is not playing allow the
performer time to listen and understand. For systems where extra sounds
are generated by a predictive model in synchrony with the musician, the
source of these sounds must be clear. One strategy is to follow
structured performance paradigms such as jazz interaction, or
live-looping; another is to physically embody these responses in robots
or visually represented instruments.

\hypertarget{sec:future-work}{%
\subsection{Final Remarks}\label{sec:future-work}}

Prediction has clear roles in musical performance. In this work we have
shown how predictive models can fit into DMI design by complementing and
extending the cognitive prediction already used by performers. Our
review has explored the musical and creative consequences of prediction
at the instrument-, performer-, and ensemble-level. In a world where AI
and deep learning interactions are increasingly built into everyday
devices, the place of predictive models in musical interaction certainly
bears scrutiny. While DMI designs show strong use of multi-modal
sensing, highly creative processing, and artistically savvy responses,
predictive models have sometimes been under-explored. We argue that
considering machine learning models in DMIs as extensions of human
cognitive predictions helps to explain their benefits to users and
performers. Deep models, such as RNNs, are being widely explored for
music modelling, but, despite their flexibility in learning large and
low-level musical sequences, are not yet widely used in DMIs. Future
deep predictive models may be able to handle multiple types of
prediction in a DMI, with end-to-end mappings from sensors directly to
sound. To achieve these deep predictive DMIs, we challenge musical
interface designers to consider prediction as a new framework for ML in
interactive music.

\hypertarget{funding}{%
\subsection{Funding}\label{funding}}

This work was supported by The Research Council of Norway as a part of
the Engineering Predictability with Embodied Cognition (EPEC) project,
under grant agreement 240862.

\hypertarget{conflict-of-interest}{%
\subsection{Conflict of Interest}\label{conflict-of-interest}}

The authors declare that they have no conflict of interest.

\begin{landscape}
\begin{longtable}{@{}p{3cm}p{6cm}p{2.5cm}p{3cm}p{4cm}l@{}}
\caption{Predictive interactive music systems reviewed in this paper
ordered by level: instrument-level (inst.), performer-level (perf.),
  and ensemble-level (ens.). The machine-learning model used, and its
  input and output configuration are also listed.}
  \label{reviewed-dmis-table}\\

\toprule
\textbf{Title} & \textbf{Description} & \textbf{Model} & \textbf{Input} & \textbf{Output} & \textbf{Level} \\
\toprule
\endhead
GloveTalk II \cite{Fels:1998fj} & speech synthesis control & MLP & hand sensors & synthesis parameters & inst. \\ \midrule
PiaF \cite{Van:2014it} & control effects with hand gestures & GVF & Kinect & audio effect parameters & inst. \\ \midrule
BRAAHMS \cite{Yuksel:2015uo} & adaptive harmonisation using BCI & SVM & fNIRS & harmonisation parameters & inst. \\ \midrule
Emotionally intelligent piano \cite{Ben-Asher:2013aa} & control visualised colours & naive Bayes & IMU & emotion classes & inst. \\ \midrule
The Birl \cite{jsnyder1:2014} & control synth through button interface & MLP & capacitive button sensors & continuous pitch & inst. \\ \midrule
Self-Supervising Machine \cite{Smith:2011aa} & control synth through unsupervised control interface & ART, MLP & touchscreen control data & synthesis parameters & inst. \\ \midrule
000000Swan's Monster \cite{Schedel:2011yg} & control extra synth and video layers & MLP, classifier & Kinect, k-bow & synthesis and video parameters & inst. \\ \midrule
AI Duet \cite{Mann:2016aa} & continue performance & RNN & symbolic music & symbolic music & perf. \\ \midrule
RoboJam \cite{Martin:2018ag} & continue performance & RNN & touchscreen control data & touchscreen control data & perf. \\ \midrule
Continuator \cite{Pachet:2003wd} & continue performing in user style & Markov & symbolic music & symbolic music & perf. \\ \midrule
MySong \cite{Simon:2008dp} & automatic accompaniment generation for vocal melodies & HMM & audio & chord sequence & perf. \\ \midrule
OMax \cite{Assayag:2006aa} & improvisation agent & Markov, FO & symbolic music & symbolic music & perf. \\ \midrule
Beatback \cite{Hawryshkewich:2010aa} & continue or fill-in drum performances & Markov & symbolic music & symbolic music & perf. \\ \midrule
GenJam \cite{Biles:2007aa} & band accompaniment in real-time & GA & symbolic music & symbolic music & ens. \\ \midrule
MalLo \cite{Jin:2015aa} & predict percussion strokes & quad. regr. & camera / Leap motion & percussion stroke time & ens. \\ \midrule
Metatone Classifier \cite{Martin:2015jk} & update interface during free improvisation & RF, Markov & control data & improvisation events & ens. \\ \midrule
Neural Touchscreen Ensemble \cite{Martin:2017ae} & accompany improvisation in real-time & RNN, RF & control data & ensemble gesture classes & ens. \\ \midrule
Orchestra in a box \cite{Raphael2003} & score-following system & HMM, Bayes'n net & audio & score locations & ens. \\ \midrule
TablaNet \cite{Sarkar:2007aa} & tabla stroke recognition and phrase prediction & kNN, Bayes'n net & audio & symbolic music & ens. \\ \midrule
Shimon \cite{Bretan:2012aa} & robotic marimba player & Markov, curve matching, & kinect, symbolic music & physical movements & ens. \\ \midrule
SHEILA \cite{Tidemann:2009aa} & robotic drum player & ESN & drum pattern class & symbolic music, motor control & ens. \\
\bottomrule
\end{longtable}
\end{landscape}

\renewcommand\refname{References}
\bibliography{references.bib}
\end{document}